\documentclass{iopart}             
\usepackage[T1]{fontenc}
\usepackage{graphicx}
\usepackage{url,graphics,epsfig}
\usepackage{amssymb}

\begin{document}
\setlength{\arraycolsep}{2.5pt}             
\jl{2}
%
%
%
\def\etal{{\it et al~}}
\def\newblock{\hskip .11em plus .33em minus .07em}
%
%
%
%
%
\setlength{\arraycolsep}{2.5pt}             

\title[Valence-shell photoionization of  Xe$^{7+}$  ions] {Valence-shell photoionization
	of Ag-like Xe$^{7+}$ ions  :  experiment and theory}

\author{A M\"{u}ller$^1\footnote[1]{Corresponding author, E-mail: Alfred.Mueller@iamp.physik.uni-giessen.de}$,
	    S Schippers$^1$,  D Esteves-Macaluso$^{2,3}\footnote[2]{Present address:
	      Department of Physics and Astronomy,
	    University of Montana, Missoula, Montana 59812, USA}$,\\
  	    M Habibi$^{3}$, A Aguilar$^{2}$,
	     A L D Kilcoyne$^2$, R A Phaneuf$^3$,\\  C P Ballance$^4$
              and B M McLaughlin$^{5,6}\footnote[3]{Corresponding author, E-mail: b.mclaughlin@qub.ac.uk}$~}

\address{$^1$Institut f\"{u}r Atom- ~und Molek\"{u}lphysik,
                         Justus-Liebig-Universit\"{a}t Giessen, 35392 Giessen, Germany}

\address{$^2$Advanced Light Source, Lawrence Berkeley National Laboratory,
                          Berkeley, California 94720, USA}

\address{$^3$Department of Physics, University of Nevada,
                          Reno, NV 89557, USA}

\address{$^4$Department of Physics,  206 Allison Laboratory,
                            Auburn University, Auburn, AL 36849, USA}

\address{$^5$Centre for Theoretical Atomic, Molecular and Optical Physics (CTAMOP),
                          School of Mathematics and Physics, The David Bates Building, 7 College Park,
                          Queen's University Belfast, Belfast BT7 1NN, UK}

\address{$^6$Institute for Theoretical Atomic and Molecular Physics,
                          Harvard Smithsonian Center for Astrophysics, MS-14,
                          Cambridge, MA 02138, USA}
%
%
\begin{abstract}
We report on  experimental and theoretical results for the photoionization of
Ag-like xenon ions, Xe$^{7+}$,  in the photon energy range 95 to 145~eV.
The measurements were carried out at the Advanced Light Source at
an energy resolution of $\Delta$E = 65 meV with additional measurements made at $\Delta$E = 28 meV
and 39 meV.  Small resonance features  below the ground-state ionization threshold,
at about 106 eV, are due to the presence  of metastable Xe$^{7+} (4d^{10} 4f~^2{\rm F}^{\circ}_{5/2,7/2})$
ions in the ion beam. On the basis of the accompanying theoretical calculations using the
Dirac Atomic R-matrix Codes (DARC), an admixture of only a few percent of metastable
ions in the parent ion beam is inferred, with almost 100\% of the parent ions
 in the $(4d^{10}5s ~^2{\rm S_{1/2}})$ ground level.
 The cross-section is dominated by  a very strong resonance associated
 with $4d \rightarrow  5f$  excitation and subsequent autoionization.
 This prominent feature in the measured spectrum is the $4d^95s5f ~^2{\rm P}^{\circ}$ resonance located
at (122.139 $\pm$ 0.01)~eV. An absolute peak cross-section of 1.2 Gigabarns
was measured at 38 meV energy resolution. The experimental natural width $\Gamma$ = 76 $\pm$ 3 meV
of this resonance compares well with the theoretical estimate of 88 meV obtained from the
DARC calculation with 249 target states.  Given the complexity of the system, overall
satisfactory agreement  between theory and experiment
is obtained for the photon energy region investigated.
\end{abstract}
%
%
\vspace{0.5cm}
\begin{flushleft}
Short title: Valence shell photoionization of Xe$^{7+}$ ions\\
\vspace{0.5cm}
\submitto{\jpb: \today}
\end{flushleft}
%
%
\maketitle
%
%
%
\section{Introduction}
Collision processes with xenon ions are of interest for UV-radiation
generation in plasma discharges~\cite{Kruecken2004a}, for fusion research~\cite{Hill1999}
and also for spacecraft propulsion~\cite{Goebel2008}.
Photoionization of Xe ions in different charge states has been investigated previously in experiments 
with Xe$^{1+}$~\cite{Sano1996,Koizumi1996,Koizumi1997,Andersen2001b,Itoh2001,Thissen2008a,Bizau2011,Schippers2014}, Xe$^{2+}$~\cite{Koizumi1997,Watanabe1998,Andersen2001b,Schippers2014}, Xe$^{3+}$~\cite{Koizumi1997,Andersen2001b,Emmons2005a,Bizau2006b,Schippers2014}, Xe$^{4+}$~\cite{Bizau2006b,Aguilar2006a,Schippers2014}, Xe$^{5+}$~\cite{Bizau2006b,Aguilar2006a,Schippers2014}, Xe$^{6+}$~\cite{Bizau2006b,Aguilar2006a}, and Xe$^{7+}$~\cite{Bizau2000}.

Here we report theoretical and experimental results for the photoionization of Ag-like Xe$^{7+}$  ions
which were measured at the Ion Photon Beam (IPB) end station of  beamline 10.0.2.\
at the Advanced Light Source (ALS) in Berkeley, California. Xe$^{7+}(4d^{10}5s ~^2{\rm S}_{1/2})$
is of particular interest because of its quasi-single-electron nature and, hence, its potential as
a test case for understanding fundamental aspects of atomic processes.
Compared with the only previous experimental study of single photoionization of
Xe$^{7+}$ carried out by Bizau and co-workers \cite{Bizau2000},
the present cross-sections were obtained at higher energy resolution
(65~meV versus more than 700 meV) and on an absolute cross-section scale.
Preliminary experimental results from the present study were reported by Schippers \etal~\cite{Schippers2009}.
In the  photon energy range of 95 - 145 eV covered by the experiment the cross section
is dominated by resonances associated with $4d \rightarrow  nf$
excitations with $n=5,6,7,...$ and subsequent autoionization.
The high-precision cross-section measurements carried
out at the ALS are compared with large-scale
theoretical calculations within the confines of the Dirac Coulomb R-matrix
approximation~\cite{venesa2012,Ballance2012,McLaughlin2012,ballance04}.
For the computations the Dirac Atomic R-matrix Codes package (DARC) was employed
using progressively larger expansions in the close-coupling calculations. The results
provide  suitable representations of photoionization of
the $(4d^{10}5s ~^2{\rm S}_{1/2})$ ground and metastable
$(4d^{10}4f ~^2{\rm F^{\circ}}_{5/2,7/2})$ levels of Ag-like Xe$^{7+}$ ions,
the latter of which are excited by about 32.9~eV~\cite{NISTasd} above the ground state.

In addition to the direct photoionization process removing the
outermost $5s$ or $4f$ electron,  the following indirect
excitation processes  can occur for the interaction of a photon with
the $5s$ ground-state  and the $4f$ metastable levels of the Xe$^{7+}$ ion:
\begin{eqnarray}
h\nu +  {\rm Xe}^{7+}(4d^{10} 5s~^2{\rm S}_{1/2}) &~ \to ~ &
   {\rm Xe}^{7+}(4d^{9} 5sn\ell ~^2{\rm L}^{\circ}_{1/2,\, 3/2} ) \nonumber \\
h\nu + {\rm Xe}^{7+}(4d^{10} 4f~^2{\rm F}^{\circ}_{5/2}) &~ \to ~ &
  {\rm Xe}^{7+}(4d^{9} 4f n{\ell}~ ^2{\rm L}_{3/2,\, 5/2,\, 7/2}) \nonumber \\
h\nu + {\rm Xe}^{7+}(4d^{10} 4f~^2{\rm F}^{\circ}_{7/2}) &~ \to ~ &
  {\rm Xe}^{7+}(4d^{9} 4f n{\ell}~ ^2{\rm L}_{5/2,\, 7/2, \,9/2} ), \nonumber
\end{eqnarray}
where the term angular momentum L is 1, 2, 3, 4 or 5 (i.e. P, D, F, G or H)
of the excited xenon ion. The contributing principal quantum numbers
$n$ (which can be as low as $n=4$ for the metastable initial state) depend
on the orbital quantum number $\ell$ (predominantly $\ell = 3$ or 1) of the
subshell to which the $4d$ electron is excited. The intermediate resonance
states can then decay by autoionization to the ground state or other
energy-accessible excited states of Xe$^{8+}$.

The layout of this paper is as follows. Section 2 details the experimental procedure used.
Section 3 gives an outline of the theoretical work. Section 4 presents and discusses the
results obtained from both the experimental and theoretical methods.
Finally in section 5 a summary is provided and conclusions are drawn from the present investigation.
%
%
%
%
%
\section{Experiment}\label{sec:exp}
A detailed description of the experimental setup has been provided
by Covington and co-workers \cite{Covington2002}. The experimental procedures
were similar to those used previously by M\"{u}ller \etal~\cite{Mueller2009,Mueller2010b}.
The Xe$^{7+}$ ions required for the present experiment were produced from natural xenon gas
inside a compact 10 GHz all-permanent-magnet electron-cyclotron-resonance ion source placed
at a positive potential of +6 kV. After extraction and acceleration of the ions to a kinetic energy of
42~keV the ion beam was deflected  by a 60$^\circ$ dipole magnet selecting ions of the desired
ratio of charge to mass. Collimated isotope-resolved $^{129}$Xe$^{7+}$ ion beams with electrical
currents of up to 15~nA were passed through the photon-ion merged-beam interaction region.
By applying appropriate voltages to several electrostatic steering and focusing devices the
ion beam  was aligned along the axis of the counter-propagating photon beam.
Downstream of the interaction region,  the ion beam was deflected out of the photon-beam direction
by a second dipole magnet that also separated the ionized Xe$\rm ^{8+}$ product
ions from the Xe$\rm ^{7+}$ parent ions. The Xe$\rm ^{8+}$ ions were counted with a single-particle
detector of nearly 100\% efficiency~\cite{Fricke1980a,Rinn1982}, and the Xe$\rm ^{7+}$ ion beam
was collected in a Faraday cup monitoring the ion current for normalization purposes.
The measured Xe$\rm ^{8+}$ count rate did not result completely from photoionization events.
Although the interaction region was maintained at ultra-high vacuum,
the product ion beam also contained Xe$\rm ^{8+}$ ions produced by electron-loss collisions with residual
gas molecules and surfaces. For the determination of absolute cross sections this background
was subtracted by employing time modulation (mechanical chopping) of the photon beam.

%
\begin{figure}
\begin{center}
\includegraphics[width=9cm]{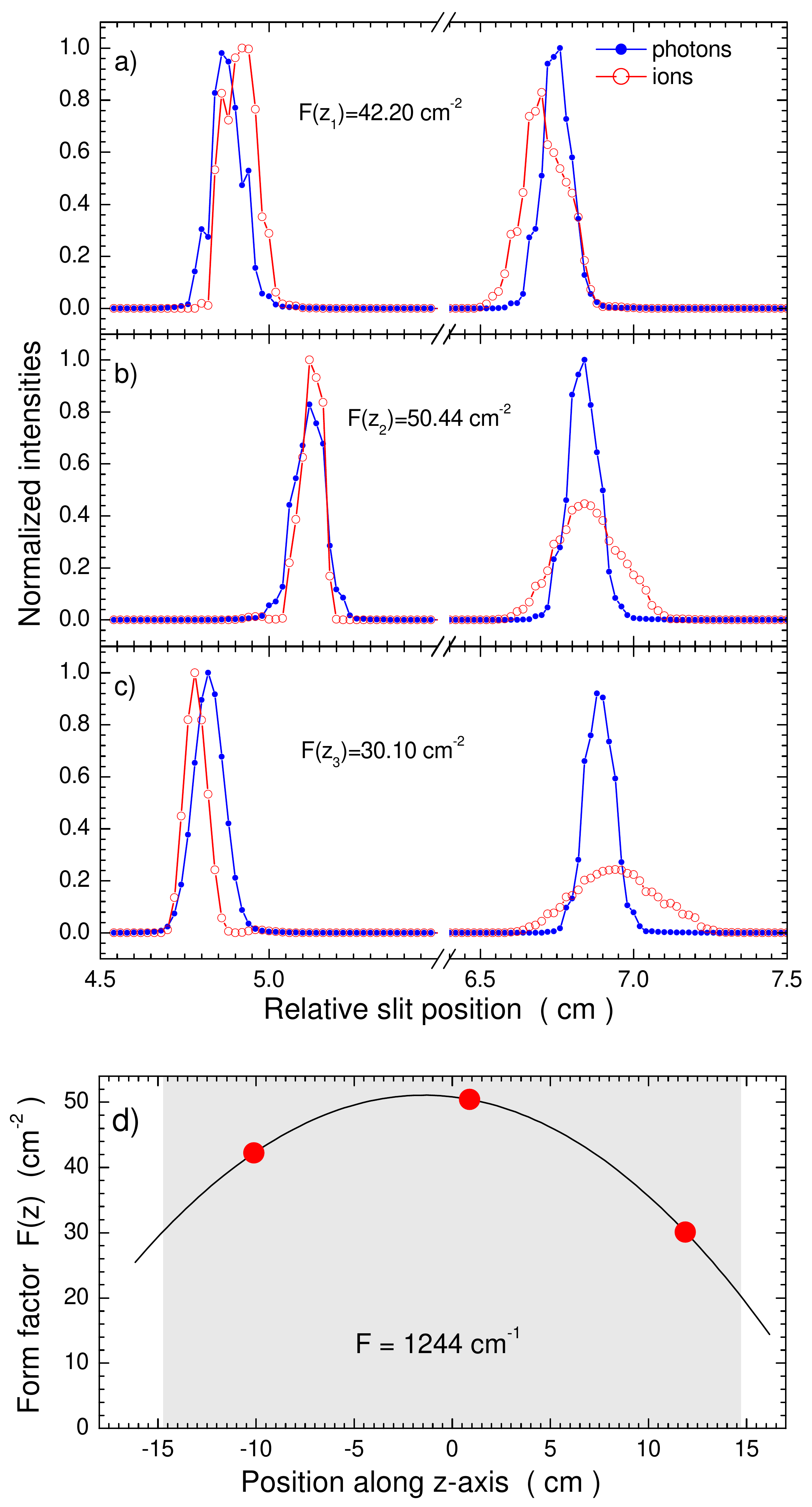}
\caption{\label{formfactor}(Colour online)  Determination of the merged-beam form factor by
						slit scans of the beam profiles at different positions z$_i$, $i$=1,2,3 along the axis of the
						overlapping photon and ion beams. Panels a), b), and c) show the measured beam
						profiles at z$_1$=-10.11~cm, z$_2$=0.88~cm and z$_3$=11.88~cm, respectively,
						where z=0 is in the center of the interaction region.
						Panel d) shows the calculated overlap integrals, F(z), at the
						measured positions (solid dots) and the solid line is a second-order polynomial fit to the
						data  used for interpolating F(z). The shaded area indicates the length of the interaction
						region. The resulting total form factor F is the integral of F(z) over the length of the
						interaction region. The numerical values of the overlaps and the total form factor
						of the specific measurement are given in the respective panels. }
\end{center}
\end{figure}

Absolute cross sections were obtained by normalizing the background-subtracted Xe$\rm ^{8+}$
photo-ion yield to the measured ion current, to the photon flux which was determined using a
calibrated photodiode, and to the geometrical beam overlap. The latter was optimized by employing
two commercial rotating-wire beam-profile monitors located in front of and behind the interaction region.
The effective length of 29.4~cm of the interaction region is defined by a drift tube which, in the
present experiment, was  set to a potential of -1.5~kV with respect to the surrounding grounded
electrodes and the vacuum vessel. By this potential, the Xe$\rm ^{8+}$ ions produced inside
the drift tube are energy-tagged. They emerge with a kinetic energy of
42~keV+7e$\times$1.5~kV-8e$\times$1.5~kV = 40.5~keV, where $7 e$ is the electrical
charge of the entering ions and $8e$ is the electrical charge of the exiting ions. Xe$\rm ^{8+}$ ions
born outside the interaction region maintain their original energy of 42~keV.  The demerging
magnet and a subsequent spherical electrostatic energy analyzer
easily separates Xe$\rm ^{8+}$ ions with energies of 40.5 and 42 keV, respectively, from one another.
The overlap of the merged beams is quantified by the form factor~\cite{Phaneuf1999} which was determined
by scanning narrow slits across the two beams at three positions along the beam axes, at the entrance,
the middle and the exit of the interaction region. Due to the considerable effort required for carrying
out reliable absolute cross-section measurements, these were performed at only a
few photon energies at selected resonance maxima. By comparison of these absolute measurements 
with the photo-ion-yield spectrum a constant normalization factor was determined matching 
the yield spectrum to the absolute cross sections. The scatter in the ratios of absolute 
cross sections and associated photo-ion-yields is much smaller than the relative systematic uncertainty of each absolute measurement.

To illustrate the absolute measurements, experimental form-factor data taken at a
photon energy of 122.14~eV are shown in figure~\ref{formfactor}. For this specific measurement
the monochromator slits were set to 59.9~$\mu $m at the entrance and 176.6~$\mu $m at the
exit position resulting in a photon-energy resolution of 65~meV. The current measured at the
photodiode was 81.75~$\mu $A corresponding to a photon flux of 3.37$\cdot$10$^{13}$~s$^{-1}$.
The electrical ion current of 1.44~nA produced detector background of ($39 \pm 1$)~s$^{-1}$
mainly by collisions of Xe$\rm ^{7+}$ with residual gas particles resulting in the
production of Xe$\rm ^{8+}$ ions. The photoionization signal rate after background
subtraction was ($1910 \pm 45$)~s$^{-1}$. The cross section resulting from this particular
measurement taken with a form factor  F = 1244~cm$^{-1}$  is 965~Mb with a statistical uncertainty of 2.3\%.

 The error budget of absolute measurements at the IPB endstation has been previously
 discussed in detail by Covington \etal~\cite{Covington2011}. Different from their experiment,
 a much more efficient single-particle detector~\cite{Fricke1980a,Rinn1982} was employed in the present measurements.
 Uncertainties associated with the product ion detection including product-ion collection were substantially reduced from a
 total of about 11\% to a new total of only 5\%. The resulting total systematic error of the present
 absolute cross-section determination is estimated to be $\pm$ 19\% at 90\% confidence.

 The energy scale was calibrated by first applying a Doppler correction to account for the motion of
 the Xe$\rm ^{7+}$ parent ions. The monochromator calibration was determined
 by measurements of the photo-ions produced in photoionization of Ar gas in the vicinity of the Ar $2p_{3/2}$
 edge which was observed in second-and third-order light from the monochromator. The Ar photoabsorption
 spectrum covers characteristic and distinct features in the energy range from about 244 to 250~eV.
 Observations of these features with second-order light were made in the nominal energy range 122 to 124.5~eV
 and with third-order light in the energy range 82.1 to 82.9~eV. The resulting resonance positions were
 compared with the measurements carried out by King \etal \cite{King1977} and Ren \etal~\cite{Ren2011}.
 King \etal \cite{King1977} employed electron-energy-loss spectroscopy (EELS) with 1.5~keV electrons at
 about 70~meV resolution reaching an accuracy for the $(2p^5 ~^2{\rm P}_{3/2}) 4s$ resonance of
 10~meV at 244.39~eV, that was unsurpassed for decades until a few years ago. Ren \etal~\cite{Ren2011}
 also measured electron-energy-loss spectra with a resolution of 55~meV using 2500~eV electrons
 scattered from Ar at 0$^\circ$ and 4$^\circ$. With improved energy resolution and better statistics they
 were able to provide the same energy of 244.390~eV for the $(2p^5 ~^2{\rm P}_{3/2}) 4s$ resonance
 but with an uncertainty of only 4~meV, i.e., a factor of two and a half better than King \etal. The comparison
 of the present photoabsorption energies in second and third order with the literature values showed
 deviations between -23 meV near 82~eV and +53~meV near 122~eV. By applying a linear correction
 to the nominal photon energies of the present experiment an uncertainty of the overall energy calibration 
 of $\pm$ 10 meV is estimated. Since the dominating Xe$^{7+}$ photoionization resonance found in the 
 present experiment to be at an energy of 122.139 eV almost coincides in position with the energy 
 to be expected in the second-order observation of the Ar($(2p^5 ~^2{\rm P}_{3/2}) 4s$ resonance, 
 known to be at 122.195 eV, we conclude  that the calibrated energy scale has a conservatively 
 estimated uncertainty of better than $\pm $ 5 meV  at the position of the Xe$^{7+}(4d^9 5s5f ~^2P^o)$ resonance.

%
%
%
%
%
\section{Theory}\label{sec:Theory}

\subsection{DARC calculations for Ag-like Xenon }
 For comparison with high-resolution measurements such as those carried out in the present study,
 state-of-the-art theoretical methods using highly correlated wavefunctions  are required that include relativistic effects
 and thus provide quantitative results on fine-structure level splitting.  An efficient parallel version \cite{ballance06}  of the
 DARC~\cite{norrington87,norrington91,norrington04,darc} suite   of codes has been
 developed \cite{venesa2012,Ballance2012,McLaughlin2012}   to address the challenge
 of electron and photon interactions with atomic systems supporting the inclusion of hundreds of levels
 and thousands of scattering channels. Metastable states  are populated in the Xe$^{7+}$ ion
 beam experiments requiring additional theoretical calculations. Recent modifications to the
 DARC codes~\cite{venesa2012,McLaughlin2012,Ballance2012} facilitate high quality
 photoionization cross-section calculations for heavy complex systems of interest to
 astrophysics and other plasma applications.
 Previously, cross-section calculations for  single
 photoionization of trans-Fe elements such as Se$^{+}$, Xe$^{+}$ and Kr$^{+}$
  ions~\cite{Ballance2012,McLaughlin2012,Hino2012} have been made
  using this modified DARC code illustrating excellent
  agreement with the available experimental measurements.

Ag-like Xe$^{7+}$ is a quasi--one-electron system with a single $4s$ electron outside filled
sub-shells in the ground level or a single $4f$ electron in the first metastable excited level
that contributes to the experimental cross section. Complexity immediately emerges when
a $4d$ electron is excited producing atomic configurations with typically 3 open sub-shells
with relatively high angular momenta. Photoionization cross-section calculations for Xe$^{7+}$
were performed for the ground ($4d^{10}5s$) and the excited metastable ($4d^{10}4f$)  levels
with a progressively larger number of states in our close-coupling work to benchmark our theoretical results
with the present high resolution experimental measurements.

The atomic structure calculations were carried out using the GRASP code \cite{dyall89,grant06,grant07}.
 Initial atomic structure calculations were done for the target states using 249-levels arising from
 twelve configurations of the Pd-like (Xe$^{8+}$) residual ion. The configurations $4d^{10}$,
 $4d^{9}4f$, $4d^{9}5s$, $4d^{9}5p$,  $4d^{9}5d$,  $4d^{9}5f$,  $4d^{9}6s$, $4d^{9}6p$,
 $4d^{9}6d$,  $4d^{9}6f$,  $4d^{8}5s^2$ and $4d^{8}5p^2$  were included in the close-coupling
 calculations. We investigated the convergence of this model by incorporating the additional
 configuration $4d^{8}5d^2$ which gave rise to a total of 526-levels. Models were also investigated
 where we left out the $6d$ orbital in the ground and metastable states and incorporated a $4f^2$ configuration.
 Photoionization cross-section  calculations were then performed using the DARC codes for the different scattering models
  for the ground $4d^{10}5s~^2S_{1/2}$ state of the Xe$^{7+}$ ion
  (with 249, 508 and 526 levels) in order to gauge convergence of the models.
 The DARC photoionization cross-section calculations for the metastable $(4d^{10} 4f)$ levels were restricted to a 528-level model.

The R-matrix boundary radius of 12.03 Bohr radii  was sufficient to envelop the radial extent of
all the n=6 atomic orbitals of the Xe$^{7+}$ ion. A basis of 16 continuum orbitals was sufficient to
span the incident experimental photon energy range from threshold  up to 150 eV. Since dipole
selection rules apply,  total ground-state photoionization requires only  the bound-free dipole matrices,
$\rm 2J^{\pi}=1^{e} \rightarrow ~2J^{\pi}=1^{\circ},3^{\circ}$. For the excited metastable
states,  $\rm 2J^{\pi}=5^{o} \rightarrow~2J^{\pi}=3^{e}, 5^{e}, 7^{e}$ and
 $\rm 2J^{\pi}=7^{o} \rightarrow~2J^{\pi}=5^{e}, 7^{e}, 9^{e}$ are necessary.

 For the ground and metastable initial states, the outer region electron-ion collision problem
 was solved (in the resonance region below and between all thresholds)
 using a suitably chosen fine energy mesh of 1.5$\times$10$^{-8}$
 Rydbergs ($\approx$ 0.2 $\mu$eV) to fully resolve all the extremely narrow resonance
 structure in the photoionization cross sections. The $jj$-coupled Hamiltonian diagonal matrices
 were adjusted so that the theoretical term energies matched the recommended
 NIST values~\cite{NISTasd}. We note that this energy adjustment ensures better positioning
 of certain resonances relative to all thresholds included in the calculation \cite{McLaughlin2012, Ballance2012}.
 The DARC photoionization cross sections for Ag-like xenon ions were convoluted with
 a Gaussian of 65~meV full-width-at-half-maximum (FWHM) in order to simulate the measurements.
 For a direct comparison with the ALS measurements an appropriate weighting of the cross sections from
 the ground and metastable state was made due to the presence of metastable states in the Xe$^{7+}$ ion beam.

\subsection{Assignment of the dominant resonance features}
\label{LANL}
The DARC results do not immediately provide detailed information about the intermediate excited
states populated during the photoionization process unless a detailed resonance analysis is carried out.
In order to gain some physical insight into the features observed in the spectra,
separate calculations for the $4d^{10}5s$  ground level and the metastable $4d^{10}4f$ levels were carried out
with the Cowan code~\cite{Cowan1981} using the Los Alamos package of atomic physics codes~\cite{LANLcode}.
Fine-structure mode was applied so that level-to-level transitions could be explored. For the $4d^{10}5s$
ground level nine 4d-excited configurations $4d^{9}5s5p$, $4d^{9}5s6p$, $4d^{9}5s7p$, $4d^{9}5s8p$,
$4d^{9}5s9p$, $4d^{9}5s5f$, $4d^{9}5s6f$, $4d^{9}5s7f$, and $4d^{9}5s8f$ were considered giving
rise to 272 energy levels. In a second step the oscillator strengths $f$ of all dipole-allowed transitions
from the ground level to excited levels within the above configurations were calculated. The resulting
oscillator strengths were used to generate cross sections for photoabsorption. These are identical to
the photoionization cross sections for each  resonance, provided the branching ratios for single
autoionization are all unity.  Under this condition the photoionization cross section can be derived as
\begin{eqnarray}
 \sigma_{\rm PI}(E)   &=& 2 \pi^2 \alpha  a_0^2 ~ R_{\infty}     \frac{df}{dE} \nonumber  \\
                     &= & 1.097618 \times 10 ^{-16}                 \frac{df}{dE}  \quad       {\rm eV ~ cm^2}
 \end{eqnarray}
where $\alpha$ is the fine structure constant, $a_0$ is the Bohr radius, $R_{\infty}$ is the Rydberg constant
and $df/dE$ is the differential oscillator strength per unit energy~\cite{Cowan1981,Fano1968}.
Rearranging and integrating the equation relates the calculated oscillator strength $f$ with the resonance strength $\overline{\sigma}^{\rm PI}$   (in Mb eV)
\begin{equation}
 f =9.11 \times 10^{-3} \int_{-\infty}^{\infty} \sigma_{\rm PI} (E) {\rm d} E =9.11 \times 10^{-3}  \quad \overline{\sigma}^{\rm PI}  \label{osc}
 \end{equation}
where $\sigma _{\rm PI}(E)$ is the photoionization cross section in Mb (1 Mb = 1.0 $\times$ 10$^{-18}$ cm$^2$),
and $E$ is the photon energy in eV. The resulting cross section $\sigma_{\rm PI}(E) = \overline{\sigma}^{\rm PI} \delta(E-E_{res})$,
(where $\delta(x-x^{\prime})$ is the standard Dirac $\delta$  function \cite{Abramowitz1972} for a continuous variable)  assumed to contribute only
at resonance energy $E_{res}$ was convoluted with a Gaussian of 65~meV FWHM in
order to simulate the experimental resonance widths. Further manipulations of the
calculated photoionization spectrum are discussed in the results section.

Calculations with the Cowan code were also carried out for the metastable $4d^{10}4f$ levels of  Xe$^{7+}$. Ten excited configurations
$4d^{9}4f5p$, $4d^{9}4f6p$, $4d^{9}4f7p$, $4d^{9}4f8p$, $4d^{9}4f9p$, $4d^{9}4f^2$, $4d^{9}4f5f$, $4d^{9}4f6f$, $4d^{9}4f7f$,
and $4d^{9}4f8f$ were included which give rise to 1570 levels including the parent $^2$F levels.
A total number of 1520 oscillator strengths $f$ of all dipole-allowed transitions from the two $^2$F$_{5/2}$ and
$^2$F$_{7/2}$ ground levels to excited levels within the above configurations were calculated and cross sections determined.

%
%
\begin{figure}
\begin{center}
\includegraphics[width=10cm]{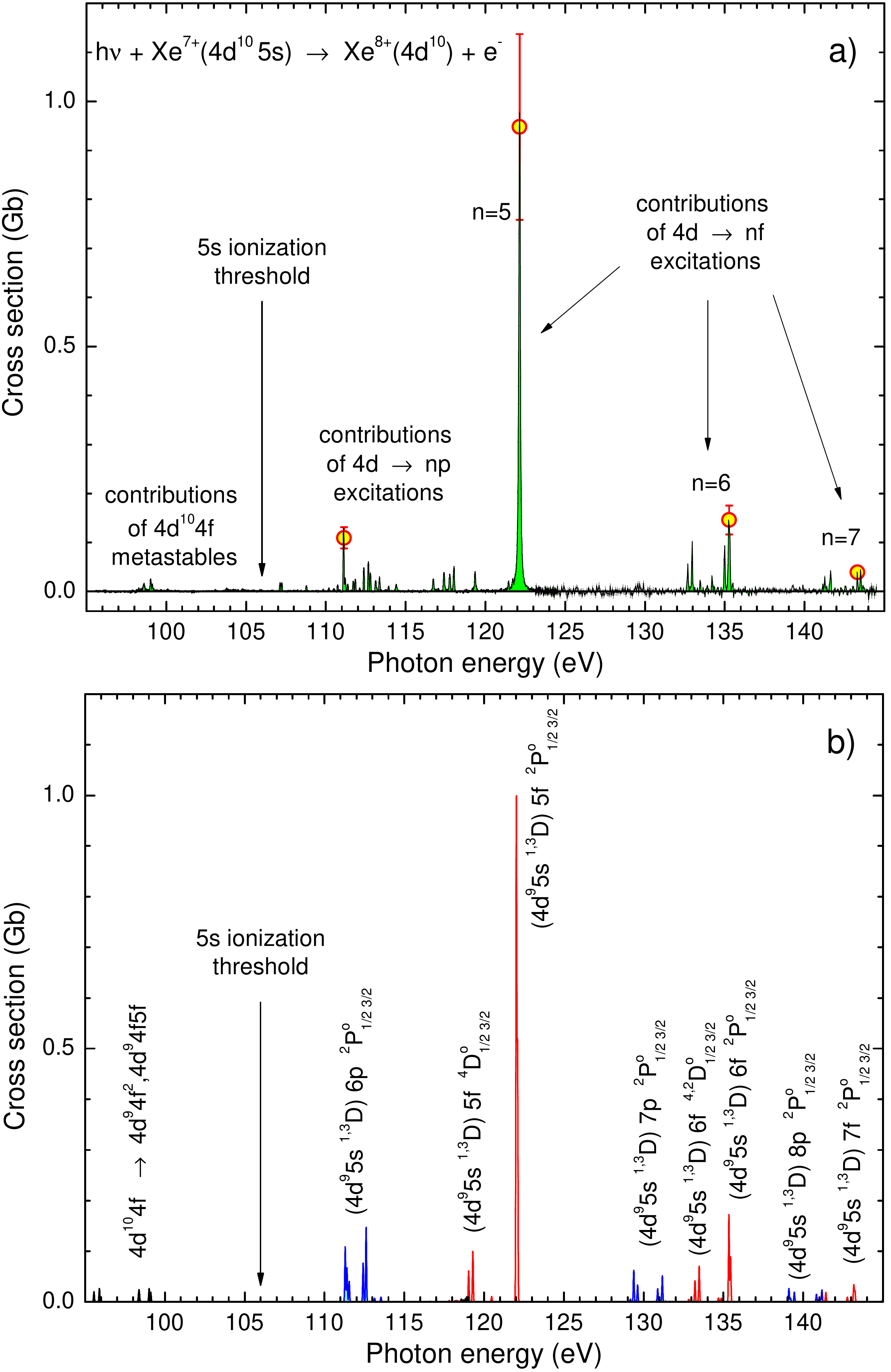}
\caption{\label{overview}(Colour online)  Overview of the valence-shell photoionization
						cross section of  Xe$^{7+}$ ions investigated in the photon
						energy range 95 - 145 eV. The measurements were made with
						a 65~meV bandwidth. Panel a) displays absolute measurements
						(large open circles with light (yellow online) shading and with
						19\% total error bars) and detailed energy-scan data
						(small solid dots with statistical error bars) connected by a solid line.
						The area under the solid line is shaded (green online).
						The ionization threshold 105.978~eV~\cite{NISTasd} of the
						ground state is indicated by the vertical arrow.
						The labels in panel a) are partly taken from panel b) which
						shows results of  relatively simple
						model calculations on the basis of the Cowan code~\cite{Cowan1981}.
						These calculations were carried out to identify the features found in
						the experiment. Resulting assignments are given in the figure, see text for details.}
\end{center}
\end{figure}
%
%
%
%
\begin{figure}
\begin{center}
\includegraphics[width=\textwidth]{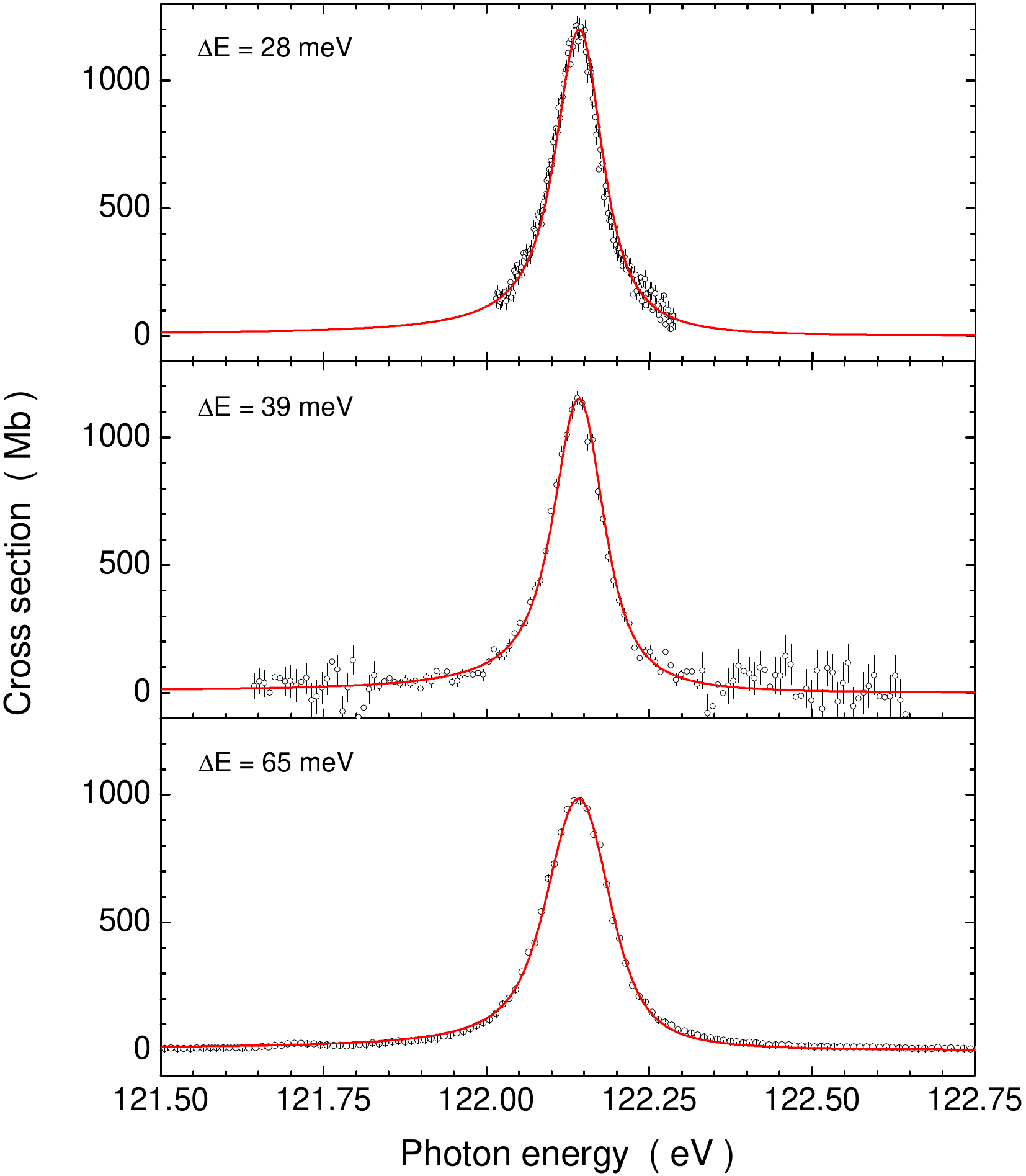}
\caption{\label{bigresonance}(Colour online)  Analysis of the dominating $4d^{9}5s5f~{\rm ^2P^o}$
						resonance formed during photoionization of ground-state  Xe$^{7+}$ ions.
						The 3 panels of this figure show measurements of the energy dependence
						of the resonance cross section taken at different bandwidths of the photon beam,
						namely, 28~meV, 39~meV and 65~meV. The solid lines result from a global
						simultaneous fit of all three measurements; see text for details.}
\end{center}
\end{figure}
%
\section{Results and Discussion}
 An overview of the experimental cross section for single photoionization of Xe$\rm ^{7+}$ ions is given in
 figure~\ref{overview}. The whole spectrum  in the photon energy range  95 - 145 eV is dominated by a
 huge resonance at approximately 122~eV. It is associated with a $4d \rightarrow 5f$ transition in ground-state Xe$\rm ^{7+}$.

A plot of  the energy dependence of $\sigma _{\rm PI}(E)$ obtained with the Cowan code (see Section~\ref{LANL})
shows many similarities with the experimental spectrum. In the calculations one can recognize the main
features obtained in the experiment, however, the resonance energies for $4d^{10}5s \to 4d^{9}5snf$ transitions
are not exactly matched. A linear transformation $E_{obs} = 9.081393 eV + 0.936797 * E_{Cowan}$ 
of the energies $E_{Cowan}$ calculated with the Cowan code and
involving shifts of the order of 1 to 2 eV or less, nicely line up the calculated resonance positions $E_{Cowan}$ with those of the
experiment, $E_{obs}$, with deviations around an average of 0.25~eV.  
The $4d^{10}5s \to 4d^{9}5snp$ transitions do not require any shifts. Also for the
$4d^{10}4f \to 4d^{9}4fnl$ transitions there is no obvious necessity to recalibrate the resonance energies.
The relative sizes of the 3 groups of resonances discussed above were adjusted to the experiment.
The result of this analysis is shown in (panel b) of figure~\ref{overview} where assignments of calculated
transitions to the dominant features are provided. The model (panel b) agrees remarkably well with the
measured spectrum (panel a). This agreement provides confidence in the given assignments and their
application to the resonance peaks observed in the experiment. Obviously, the dominant resonance is
associated with $4d^{10}5s ~^2{\rm S}_{1/2} \to 4d^{9}5s5f ~^2{\rm P^{\circ}}_{1/2,\, 3/2}$ transitions. Next in
size are resonances associated with $4d^{10}5s ~^2{\rm S}_{1/2} \to 4d^{9}5s6f ~^2{\rm P^{\circ}}_{1/2,\, 3/2}$
and $4d^{10}5s ~^2{\rm S}_{1/2} \to 4d^{9}5s7f ~^2{\rm P^{\circ}}_{1/2,\, 3/2}$ transitions. However, with increasing
principal quantum number $n$, the resonance heights also decrease. This is even more pronounced in
the sequence of $4d^{10}5s \to 4d^{9}5snp$ transitions where the experiment does not show
much evidence for $n\geq7$ contributions.

The huge resonance due to $4d^{9}5s5f~^2{\rm P^{\circ}}_{1/2,\, 3/2}$ intermediate levels in the photoionization
of ground-state Xe$^{7+}$ ions reaches a peak cross section of approximately 1~Gb at a photon energy resolution
of 65~meV. Because of its relative importance and overwhelming dominance in the spectrum it has been
investigated in special detail. Figure~\ref{bigresonance} shows three separate measurements of this
resonance at energy resolutions 28~meV, 39~meV and 65~meV. None of these measurements provides 
direct evidence for a significant fine structure splitting of the $4d^{9}5s5f~^2{\rm P^{\circ}}_{1/2,\, 3/2}$ levels. 
The theoretical calculations indicate the splitting between the $4d^{9}5s5f~^2{\rm P^{\circ}}_{1/2,\, 3/2}$ levels 
 is $\sim$ 15 meV which we determined from Fano-Beutler fits to cross section calculations for the individual 
 $1/2$ and $3/2$ odd scattering symmetries. Since the resonances sit on top of each other no visible splitting 
 is seen in the total cross section.  As the splitting is much smaller than the natural widths of the two components 
we therefore fitted the data only with one-resonance Voigt profiles. The solid line in the figure represents
the result of a global fit of Gaussian-convoluted Fano profiles~\cite{Fano1968,Schippers2011} to all three
measurements simultaneously using the natural width and the Fano asymmetry parameter $q$ of the
resonance as global fit parameters. The result of this analysis is a Lorentzian width
$\Gamma$~=~76~$\pm$~3~meV of the huge resonance. The asymmetry of the profile is small.
The fit results in an asymmetry parameter $q$ = -78 $\pm $ 15. The $4d^{9}5s5f ~^2{\rm P^{\circ}}_{1/2,\, 3/2}$
resonance energy is (122.139 $\pm$ 0.005)~eV and the line strength is (161 $\pm $ 31)~Mb eV.
According to equation~\ref{osc} this corresponds to an absorption oscillator strength $f$ = 1.47 $\pm $ 0.28
which is more than 1/7 of the total absorption oscillator strength of all the 10 electrons in the $4d$ subshell.
The dominance of the resonance in the experimental spectrum is illustrated by its relative contribution
 of about 40\% to the total  resonance strength observed in the experiment. This is in respectable
 agreement with the 47\% contribution of the dominant resonance to the
 theoretical (DARC) spectrum in the same energy range.
%
\begin{figure}
\begin{center}
\includegraphics[width=\textwidth]{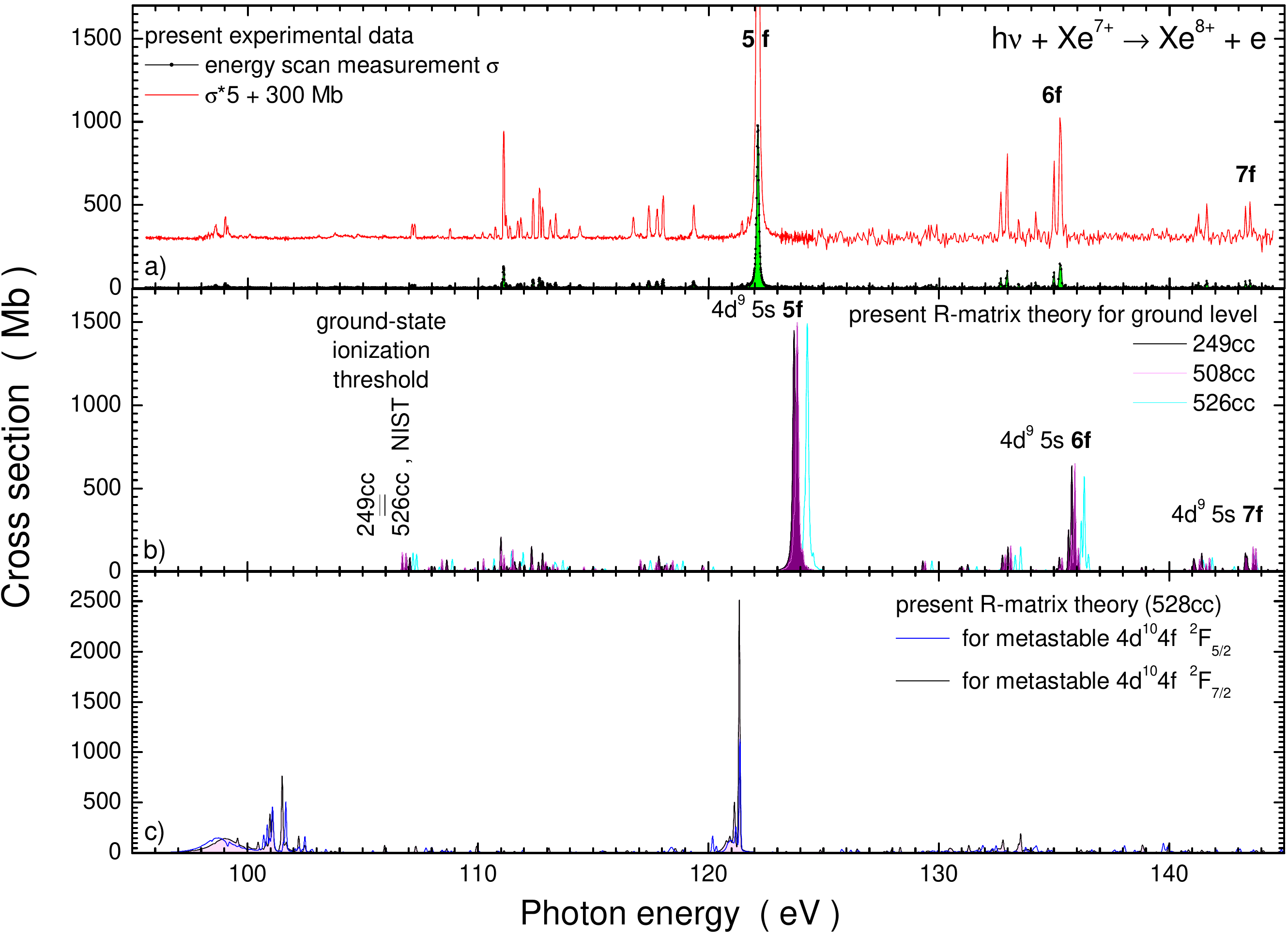}
\caption{\label{overview2}(Colour online)  The experimental and theoretical results
							of the present study of photoionization of Xe$^{7+}$ ions.
							The top (panel a) displays the experimental energy-scan data
							from figure~\ref{overview} which were taken at 65~meV resolution
							(data points connected by straight lines with light shading).
							In order to visualize the contributions of resonances other than
							the dominating $4d^9 5s 5f ~^2{\rm P}$ peak, the experimental data
							were multiplied by a factor of 5 and offset by 300~Mb (solid line, red online).
							Panel b) displays the present DARC results for a 249-level, a 508-level and
							a 526-level calculation, abbreviated as 249cc, 508cc and 526cc,
							respectively, for photoionization of ground-state Xe$^{7+}$ ions.
							 Panel c) shows 528-level DARC results for the two metastable
							 fine-structure components of the long lived excited
							 $4d^{10}4f ~^2{\rm F}$ term. The theoretical cross sections
							 were convoluted with a 65~meV FWHM Gaussian in order to
							 be comparable with the experiment.}
\end{center}
\end{figure}
%

%
\begin{figure}
\begin{center}
\includegraphics[width=\textwidth]{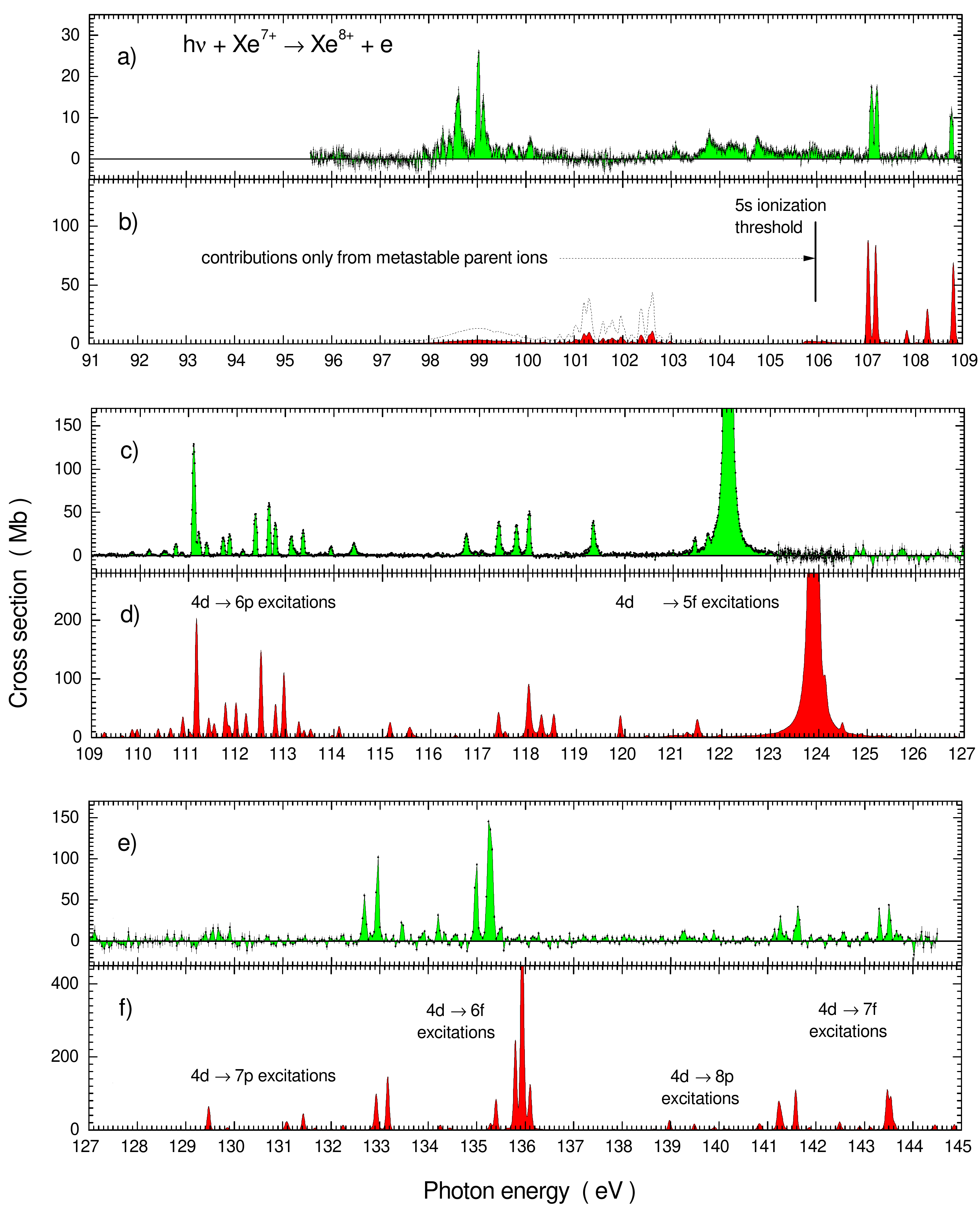}
\caption{\label{comparison}(Colour online)  Comparison of the present absolute experimental
							cross section for valence shell photoionization
							of Xe$^{7+}$ ions in the 95 - 145 eV photon energy range with the
							present 249-level DARC calculations assuming a 2.4\% metastable
							fraction in the parent ion beam and population of the fine-structure
							components according to statistical weight. Panels are a) experiment, b) theory,
							c) experiment, d) theory, e) experiment and f) theory. The cross section axis is
							chosen so that the smaller resonances apart from the overwhelming
							$4d^9 5s 5f ~^2{\rm P}$ peak can be clearly seen. It should be noted
							that the cross section axis for the theoretical cross section is different
							from that of the experimental data indicating differences in the absolute
							heights of resonance features. Nevertheless, resonance groups seen
							in the experiment can be clearly recognized in the theoretical spectrum
							as well, with reasonable agreement in relative peak heights. In order to
							better visualize the contribution of metastable Xe$^{7+}$ the associated
							cross section has been multiplied by a factor 4 and is displayed again
							as a dashed line. The main groups of $4d$-subshell excitations are indicated.}
\end{center}
\end{figure}
%

The photoionization cross section results of the experiment and of the Dirac Coulomb R-matrix calculations
for the Xe$^{7+}(4d^{10}5s~^2{\rm S}_{1/2}$) ground level as well as the metastable
Xe$^{7+}(4d^{10} 4f~ ^2{\rm F}^o_{5/2})$ and Xe$^{7+}(4d^{10} 4f~ ^2{\rm F}^o_{7/2})$ levels are
provided in figure~\ref{overview2}. The top (panel a) shows the experimental energy-scan results from
figure~\ref{overview} again with light (green online) shading and individual data points connected by
straight lines. Because of the overwhelming dominance of the resonance at 122.14~eV the much smaller
other resonances are barely visible on this scale. In order to visualize these more clearly, we have scaled them up in size.
Panel a) shows the same cross section multiplied
by a factor of 5 with the zero line shifted up by 300~Mb.
The resulting function is represented by the solid line (red online).
The middle panel b) displays the theoretical results for the Xe$^{7+}$ ground level.

In order to check the convergence of the theoretical approach  249-level, 508-level and 526-level DARC
photoionization cross-section calculations were performed for the Xe$^{7+}(4d^{10}5s~^2{\rm S}_{1/2}$)
ground state. Panel b) of figure~\ref{overview2} therefore includes the theoretical cross-section results
from all three multi-state models. The theoretical cross sections were convoluted with a 65~meV FWHM Gaussian.
Obviously, the 526-level, the 508-level and the 249-level approximations yield similar results. Unexpectedly, though,
 the 526-level calculations are slightly shifted in energy from the experiment. Thus, the 249-state model
 gives marginally better agreement with experiment. Although the experimental data were obtained with
 a mixture of ground-level and metastable ions and the theoretical results displayed in panel b) are
 only for ground-level ions, one can see here already that the DARC calculations overestimate the
 contribution of the $4d^9 5s 6f$ resonance features relative to the strong $4d^9 5s 5f$ peak.

The lowest (panel c) of figure~\ref{overview2} shows the results of 528-level DARC calculations for
the metastable Xe$^{7+}(4d^{10} 4f~ ^2{\rm F}^o_{5/2})$ and Xe$^{7+}(4d^{10} 4f~ ^2{\rm F}^o_{7/2})$ levels.
At least as far as the dominant resonance features are concerned the two spectra are very similar to one another.
According to the calculations with the Cowan code and on the basis of the assignments given in figure~\ref{overview}
the lower-energy features are due to $4d^{10} 4f \to 4d^{9} 4f^2$ excitations with possible contributions from states
within the $4d^{9} 4f 5f$ configuration. Due to the presence of 3 open subshells in this configuration, all with high
angular momenta, a  high number of possible couplings and associated energy levels are to be expected.
The numerous resulting resonances are spread out over a very broad energy range. Nevertheless a very
distinct feature can again be associated with the $4d \to 5f$ excitation of metastable Xe$^{7+}$ which theory
finds at around 121~eV close to where the huge resonance due to $4d^{9}5s5f ~^2{\rm P^{\circ}}_{1/2,\, 3/2}$
intermediate levels in the photoionization of ground-state Xe$^{7+}$ ions is located. In comparison with
the theoretical data for photoionization of metastable ions the experiment shows very small cross sections.
This is particularly true for the region below the ground-state ionization threshold which is indicated by
vertical bars in (panel b) of figure~\ref{overview2}. In this discussion one has to keep in mind that the experimental cross section is determined under the a-priori assumption of only one primary-beam component. A posteriori, the discussion shows that a small fraction $a$  of metastable ions accompanied the large fraction $(1-a)$ of ground-state ions in the parent ion beam. So, in fact the apparent experimental cross section is that for a mixture of ionic levels. The real cross section for metastable ions can in principle be obtained by normalizing the apparent cross section for signal arising from the metastable ions to the fraction $a$.  Thus, the present discussion addresses the size of that fraction $a$.

The total oscillator strength found in the experiment in
the energy range 95 - 105 eV is only 2.4\% of what theory finds in that same range. A comparably
small fraction $a$ of metastable ions in the parent ion beam must be concluded. For further comparison
between theory and experiment fractions of 97.6\% ground-level and 2.4\% metastable Xe$^{7+}$ ions
in the parent ion beam are assumed. The metastable $^2{\rm F}^o$ fraction is divided into 3/7 for
the $^2{\rm F}^o_{5/2}$ and 4/7 for the $^2{\rm F}^o_{7/2}$ contributions assuming level
population proportional to the statistical weights.

Figure~\ref{comparison} compares the present experimental cross sections with the initial-fractions
weighted theoretical 249-level DARC results except for the overwhelmingly dominant resonance peak.
As figure~\ref{overview2} already shows, the height of the dominating peak that results from excitation
of the Xe$^{7+}(4d^{10}5s~^2{\rm S}_{1/2}$) ground state to  Xe$^{7+}(4d^{9}5s5f ~^2{\rm P^{\circ}}_{1/2,\, 3/2})$
levels with a vacancy in the $4d$ sub-shell is overestimated by theory. Considering the $\pm$~19\% total
experimental uncertainty, the theoretical peak cross section is still about 28\% above the maximum
experimental limit. This is partly due to the fact that theory finds an additional resonance feature of unknown origin near the
position of the $^2{\rm P^{\circ}}_{1/2,\, 3/2}$ resonance peak. The additional resonance is seen in all DARC calculations using the different (249-, 508- and 526-level) models and is slightly shifted in energy in the different calculations. It interferes in the calculations with the Xe$^{7+}(4d^{9}5s5f ~^2{\rm P^{\circ}})$ resonance. Nevertheless, the line shape and natural width of the Xe$^{7+}(4d^{9}5s5f ~^2{\rm P^{\circ}})$  resonance can be recovered from a two-resonance fit of the unconvoluted theoretical cross-section curve in the energy range of interest assuming two Fano profiles.

\begin{figure}
\begin{center}
\includegraphics[width=\textwidth]{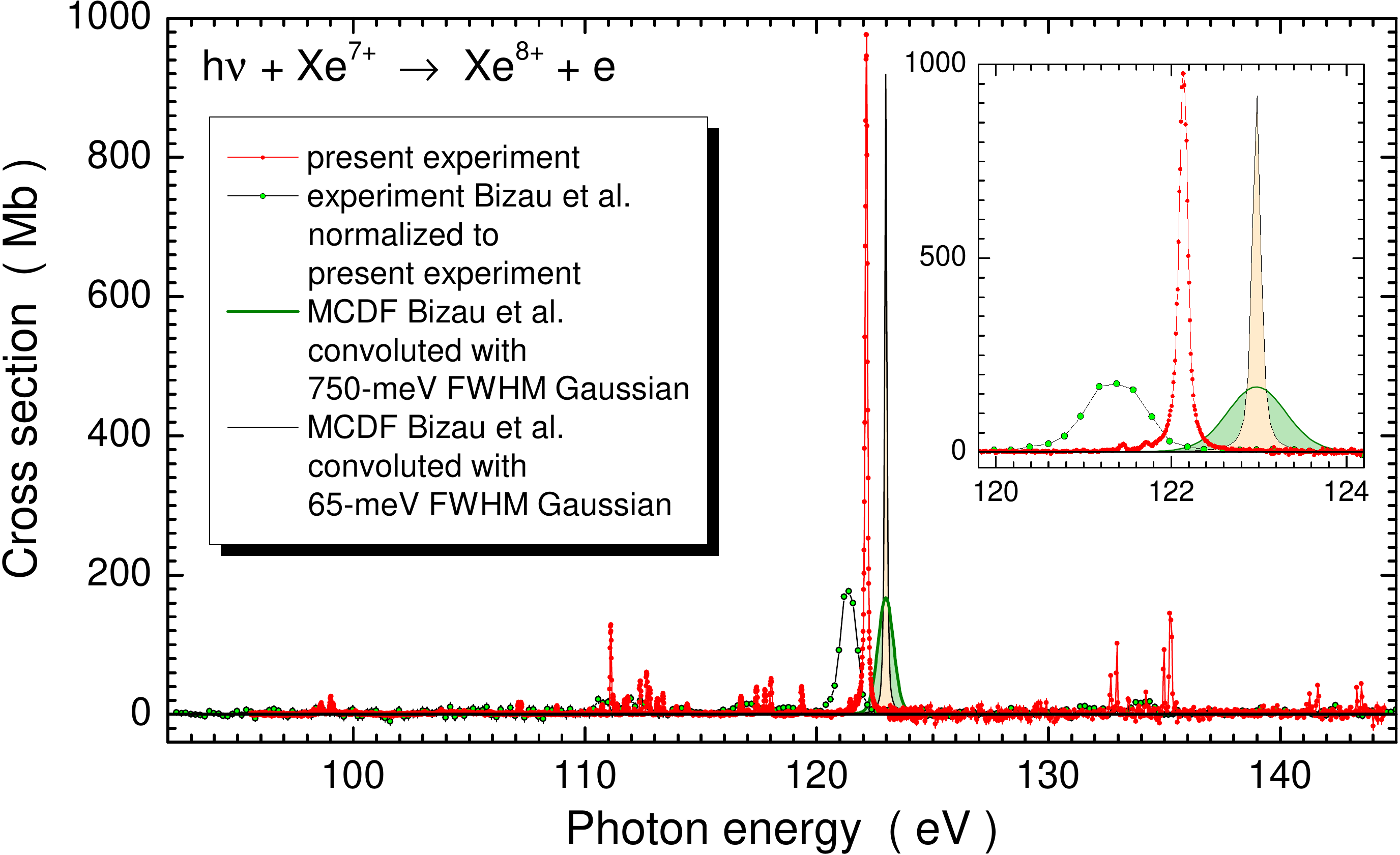}
\caption{\label{bizau}(Colour online) Comparison of the present absolute experimental
							cross section (solid red line with dots) for valence shell photoionization
							of Xe$^{7+}$ ions in the 95 - 145 eV photon energy range with previous
							experimental and theoretical work of Bizau and co-workers \cite{Bizau2000}.
							The experimental measurements made at SuperACO (solid  line with green dots)
							were taken at an energy resolution of 750 meV. The theoretical cross section
							calculations shown were obtained from the
							Multi-Configuration-Dirac-Fock (MCDF) approximation
							convoluted at 750~meV (light green shaded area),
							and  at 65~meV (light orange shaded area ).}
\end{center}
\end{figure}

In reality, the additional peak may be shifted away in
energy from the dominant peak (there is no evidence for it in the experimental data). The assumption of such a shift reduces the theoretical cross section for
photoionization involving the Xe$^{7+}(4d^{9}5s5f ~^2{\rm P^{\circ}}_{1/2,\, 3/2})$ levels by about 5.5\% bringing
it closer to the experiment. The Lorentzian width of the dominant excited $^2{\rm P^{\circ}}$ term deduced
from the theoretical data is 88~meV which compares favorably with the experimental width of (76 $\pm$ 3)~meV.
The resonance strength predicted for the $^2{\rm P^{\circ}}$ term, however, is 279~Mb~eV, more than 70\% above
the experimental result. It is worth mentioning at this point that the theoretical
Multi-Configuration Dirac-Fock (MCDF) result accompanying the previous study
of photoionization by Bizau \etal~\cite{Bizau2000} is a little more than 10\%
 below the present experimental result (see figure~\ref{bizau}).
The simple calculations carried out with the Cowan code for finding assignments of
resonances  underestimates the experimental finding by 80\%.
These examples clearly show the difficulty to predict the correct
resonance parameters for this three-open-subshells resonance feature.
%
%

It should be noted that the cross section axis in figure~\ref{comparison} for the theoretical
cross sections is different from that of the experimental data indicating differences in
absolute height of resonance features. Nevertheless, resonance groups seen in the
experiment can be clearly recognized in the theoretical spectrum with reasonable
agreement in relative peak heights. Theory and experiment agree in that they
 both show negligible contributions of direct photoionization of the outermost
 electron in Xe$^{7+}$. The cross section is dominated by inner-shell excitations
 with subsequent Auger or Coster-Kronig decays. Given the complexity of
 Ag-like Xe$^{7+}$  arising from opening the 4d sub-shell, the present study
 shows reasonable agreement of theory and experiment over the photon energy region investigated.

 Figure~\ref{bizau} provides a comparison of the present experimental data  with
 experimental and theoretical results of Bizau \etal~\cite{Bizau2000}. The inset focuses on the
 dominant Xe$^{7+}(4d^{9}5s5f ~^2{\rm P^{\circ}}_{1/2,\, 3/2})$ resonance feature. For the comparison,
 the relative results of Bizau \etal were normalized so as to match  the peak area of the
 dominant $^2{\rm P^{\circ}}$ resonance. Due to the larger width ($\approx 750$~meV instead
 of 65~meV) the normalized $^2{\rm P^{\circ}}$-peak cross section is much smaller in height
 than the present absolute measurement. For a meaningful comparison of the MCDF theory
 result with the experimental data displayed in figure~\ref{bizau} the theory curve displayed
 in Ref.~\cite{Bizau2000} for the $^2{\rm P^{\circ}}$ resonance was digitized and the resonance
 parameters were determined. With the  peak area (147~Mb~eV) and the resonance energy (122.98~eV)
 thus obtained  plus assuming a Lorentzian width of 76~meV as found in the present study, the
 MCDF resonance was convoluted with Gaussians of 65~meV and 750~meV full width at half
 maximum (FWHM), respectively, to be compared with the present and the previous experiments.
 While very good agreement is found in terms of peak height the experimental and theoretical
 resonance energies found in Ref.~\cite{Bizau2000} differ from one another and from the present
 measurement. As mentioned above, calculating the correct resonance energies in an ion with
 three open subshells is difficult. The present 249-level R-matrix calculation finds the
 Xe$^{7+}(4d^{9}5s5f ~^2{\rm P^{\circ}})$ resonance at 123.89~eV, the MCDF result is 122.98~eV
 while the present experiment finds 122.139~eV. Also the experiment by Bizau \etal with a
 resonance energy of 121.36~eV differs from the present result by 0.779~eV. While this
 difference is of the order of the experimental energy spread of the previous measurement,
 the absolute calibration could be substantially better than the resolution. However, in the
 letter format of their publication \cite{Bizau2000} Bizau~\etal did not provide an estimate
 for the uncertainty of their photon-energy axis. In contrast to that, the present experiment
 involved an extremely careful energy calibration with an estimated uncertainty of the
 Xe$^{7+}(4d^{9}5s5f ~^2{\rm P^{\circ}})$ resonance energy of only 5~meV.

\section{Summary and conclusions}\label{sec:Conclusions}
High-resolution absolute cross section measurements for single photoionization of Ag-like
Xe$^{7+}$ ions forming Xe$^{8+}$ products were obtained at the Advanced Light Source.
Compared to a previous experiment, the cross sections are now on an absolute scale and
the energy resolution and statistical precision have been vastly improved.  The experimental
results are compared with large-scale close-coupling calculations within the Dirac Coulomb
R-matrix approximation.  Calculations using different basis sets and incorporating a progressively
larger number of levels (respectively 249, 508 and 526 levels) in the close-coupling calculations,
show no significant improvement. This would indicate that convergence is already
obtained with the 249-level calculation. However, differences still exist between
theory and experiment with respect to level energies (resonance positions)
and resonance strengths. Nevertheless, the main features in the experimental spectrum are
reproduced reasonably well by theory. While the Xe$^{7+}$ parent ion in this study is a quasi-one-electron
system with a $5s$ or a $4f$ electron outside fully occupied electronic sub-shells,
the intermediate excited states (which then decay by autoionization) comprise
mainly three open sub-shells with relatively high angular momenta and the appropriate coupling of
the active electrons renders this problem quite complex.  Including many more target configurations
and levels in the close-coupling calculations makes the theoretical treatment extremely challenging.
Given the complexity of the system and the available finite computational resources, the comparison
of the present experimental high-resolution photoionization cross sections with the
large-scale close-coupling calculations in the present study shows overall agreement.

\ack
We acknowledge support by Deutsche Forschungsgemeinschaft under
project number Mu 1068/10  as well as by the US Department of Energy (DoE)
under contract DE-AC03-76SF-00098 and grant  DE-FG02-03ER15424.
C P Ballance was supported by US Department of Energy (DoE)
grants  through Auburn University.  B M McLaughlin acknowledges support by the US
National Science Foundation through a grant to ITAMP
at the Harvard-Smithsonian Center for Astrophysics,
a visiting research fellowship (VRF) from Queen's University Belfast and the hospitality of AM, SS
and the University of Giessen.  The computational work was carried out at the National Energy Research Scientific
Computing Center in Oakland, CA, USA, the Kraken XT5 facility at the National Institute
for Computational Science (NICS) in Knoxville, TN, USA
and at the High Performance Computing Center Stuttgart (HLRS) of the University of Stuttgart, Stuttgart, Germany.
The Kraken XT5 facility is a resource of the Extreme Science and Engineering Discovery Environment (XSEDE),
which is supported by National Science Foundation grant number OCI-1053575.
The Advanced Light Source is supported by the Director,
Office of Science, Office of Basic Energy Sciences,
of the US Department of Energy under Contract No. DE-AC02-05CH11231.
%
%
%
%
\section*{References}
\bibliographystyle{iopart-num}

\providecommand{\newblock}{}

\end{document}